\documentclass[prl,amsmath,aps,twocolumn,superscriptaddress]{revtex4}
\usepackage{graphicx,xspace}
\usepackage{float}
\usepackage{amsmath}
\usepackage{amssymb}
\usepackage[normalem]{ulem}
\input{epsf}
\usepackage{natbib}

\begin{document}

\title{Direct determination of the collective pinning radius 
in high temperature superconductors}

\author{M. I. Dolz, A. B. Kolton and H. Pastoriza}
\affiliation{Centro At\'omico Bariloche, Comisi\'on Nacional de Energ\'{\i}a At\'omica
R8402AGP S. C. de Bariloche, Argentina}

\begin{abstract}
We study finite-size effects at the onset of the irreversible magnetic behaviour 
of micron-sized Bi$_2$Sr$_2$CaCu$_2$O$_{8+\delta}$ single crystals by using silicon 
micro-oscillators. We find an irreversibility line appearing 
well below the thermodynamic Bragg-glass melting line 
at a magnetic field which increases both with 
increasing the sample radius and with 
decreasing the temperature. 
We show that this size-dependent irreversibility line can be identified with the 
crossover between the Larkin and the random manifold regimes of the 
vortex lattice transverse roughness. Our method 
allows to determine the three-dimensional weak collective pinning 
Larkin radius in a {\it direct way}. 
\end{abstract}
\maketitle
Superconducting vortex lattices can display the universal 
glassy properties that emerge from the 
frustrating competition between elasticity and disorder. They  
provide an exceptional test ground for the  
universal predictions of statistical 
elastic-field theories, whose methods can be indistinctly applied to periodic 
systems, such as charge density waves~\cite{gruner88,brazovskii04} and 
Wigner crystals~\cite{andrei88}, or to interfaces, such as 
magnetic~\cite{lemerle,metaxas,lee4} and ferroelectric~\cite{paruch} 
domain walls, liquid menisci~\cite{moulinet} and fractures~\cite{ponson}.

Larkin and Ovchinikov~\cite{larkin79}
demonstrated the unavoidable impact of arbitrarily weak 
disorder on the otherwise perfect vortex lattice and determined the 
basic length-scales of the problem: 
it is the finiteness of these, so-called Larkin lengths, 
what fundamentally explains the mere existence of pinning and measures 
its effective strength on the extended system~\cite{blatter94}. 
In the modern elastic theory, designed 
to correctly describe the large-scale static and dynamical 
universal behaviour of the elastic manifold, the Larkin lengths are the 
fundamental input for making quantitative  
predictions for a given experimental system.
Determining the Larkin lengths for a  
high-T$_c$ superconductor remains 
a difficult challenge however. 
Indirect empirical estimates based on 
transport properties such as the 
critical current or the creep barriers  
are an alternative, but they spoil precise 
comparisons between experiments and 
theory. Recently, 
an experimental finite-size analysis was 
applied to determine the characteristic 
dynamical length, predicted by the (bulk) 
elastic theory, controlling the 
domain wall creep motion in 
ferromagnetic nanowires~\cite{lee4}.
Here we report 
a finite-size study of the onset of irreversibility
in a micron-sized superconductor 
that allows to determine 
the Larkin radius in a 
{\em direct} way.
\begin{figure}
\includegraphics[width=0.45\textwidth]{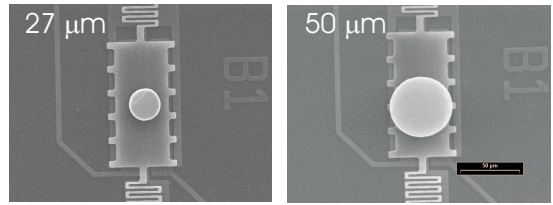}
\caption{Scanning electron microscope images of two samples 
mounted in the Silicon micro-oscillators.
Left (right) sample of radius 6.75 (25) $\mu$m.}
\label{fig:photos}
\end{figure}

The elastic theory characterizes the translational 
order by the roughness function  $W({\bf r}) = 
\langle [u({\bf r})-u({\bf 0})]^2 \rangle$, with $u({\bf r})$ 
the vortex displacement field with respect to the perfect 
lattice. For the 
Bragg-glass (BG) phase
~\cite{giamarchi95}, 
expected for weak pinning at low enough temperatures, 
a logarithmic growth of $W$ is predicted at large distances.
At short distances however, we have a {\em Larkin} regime where  
displacements grow as $W \sim r^{4-d}$, with $d=3$ the internal dimension of 
the elastic manifold. A crossover 
to the {\em Random Manifold} (RM) regime at a distance ${\bf r}_c$ 
occurs when 
displacements are comparable to the pinning force range 
$r_p$, such that $W({\bf r}_c) \sim r_p^2$. 
For a superconductor with vortices directed along the 
direction $\hat z$ of an external magnetic field, 
this crossover defines the longitudinal 
($L_c \equiv |{\bf r}_c.\hat{z}|$) 
and transverse ($R_c \equiv |{\bf r}_c- L_c \hat{z}|$) 
Larkin lengths. Besides describing a geometrical 
crossover the Larkin lengths also determine 
the size of the minimum bundle of vortices that can be 
individually pinned by the quenched disorder~\cite{larkin79}.

We fabricated 
Bi$_2$Sr$_2$CaCu$_2$O$_{8+\delta}$ samples with a procedure 
similar to that of 
Wang {\it et al.}~\cite{wang01}. Disks of radii $R_s = 6.75, 13.5,  
25\,\mu$m and $d = 1\,\mu$m of thickness 
were made-up by lithography and ion etching (see Methods) and then 
glued to high-Q silicon torsional 
micro-oscillators~\cite{bolle99,dolz08}. 
Scanning electron microscope images of two samples are shown in Fig.~1. 
When a external magnetic field is applied perpendicular to the superconducting planes 
and the torsional axis, the change in the resonant frequency $\Delta \nu_r$ 
of the oscillator is 
proportional to 
the magnetization of the sample~\cite{dolz07}.

In Fig.~2 we plot the results obtained for 
$R_s=13.5\, \mu$m
under two different protocols. In the FC protocol we cool the sample below its 
critical temperature at an applied field of $176$~Oe while registering 
$\Delta \nu_r$ (upper curve). In the ZFC protocol we cool the sample at zero field up to  
the lowest temperature, apply the same field as before, and then measure $\Delta \nu_r$ while warming up the sample (lower curve). The onset of irreversibility can be defined at the merging 
of both curves. Similar data can be obtained from fields loops at constant 
temperature as shown in the inset of Fig.~2. 
Two features can be readily observed: the clear size dependence of the 
irreversibility line, and the wide spanning in temperature of the reversible 
state compared to that of bulk 
samples~\cite{pastoriza94}. Phenomenologically, reversibility is reached when thermal fluctuations overcome the stronger pinning mechanism present in the sample. It has been argued that in this material
geometrical~\cite{majer95,zeldov94} or surface barriers~\cite{james97}  were the responsible of the irreversibility. 
Several aspects of the data point against these as the cause of the irreversibility. 
Geometrical and surface barriers decreases as the aspect ratio (thickness/diameter) increases~\cite{brandt99}. Our data shows the opposite behaviour, irreversibility is enhanced as the sample aspect ratio grows as can be directly seen 
in the inset of Fig.~2. Moreover, our data does not comply with the 
predictions given for the temperature dependence of the
geometrical barrier and its scaling with the first penetration field~\cite{brandt99}. 
It does not comply neither with the expected 
shape of the magnetization loops~\cite{campbell72} for a surface barrier. We show, 
in the following, that these puzzling finite-size effects can be explained, however
by a different and more fundamental mechanism.
\begin{figure}
\includegraphics[width=0.45\textwidth]{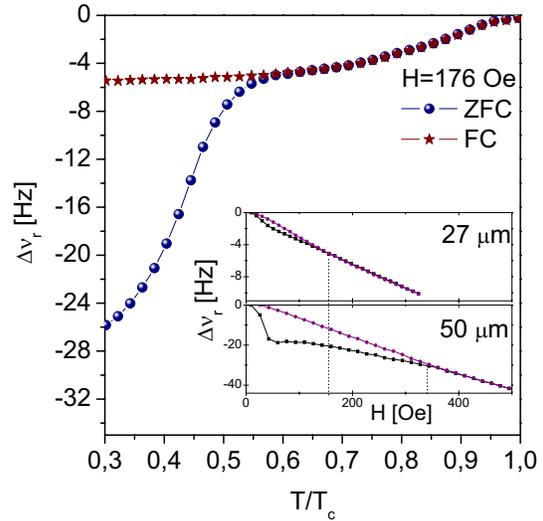}
\caption{Change in the resonant frequency of the oscillator as a function 
of temperature ($t=T/T_C$) for the sample 
of radius 13.5 $\mu$m with an magnetic field of 176 Oe applied perpendicularly to the superconducting planes in a ZFC (blue dots) and FC (red stars) experiment. 
Inset: $\Delta \nu_r$ as a function of $H$ at 
$T/T_C = 0.58$ for two sample sizes. 
These 
measurements allow the determination 
of the size-dependent 
irreversibility line. 
}
\label{fig:rawdata}
\end{figure}

We shall analyse the onset of irreversibility as the finite-size 
crossover at the vortex lattice Larkin length. 
For an applied field parallel to the $c$-axis, and 
neglecting the compression modulus 
contribution, we can use the Larkin-Ovchinikov perturbative result\cite{larkin79}, 
\begin{equation}
W({\bf r}) \approx 
r_p^2 \epsilon^4 \left[ \frac{a_0}{l_c} \right]^3 \left[ \frac{R^2}{\lambda^2} + \frac{a_0^2 L^2}{\lambda^4} \right]^{1/2},
\label{lo_nondispersive}
\end{equation}
where the so-called single-vortex collective pinning length $l_c$ absorbs the 
effective pinning strength~\cite{blatter94}, 
$a_0=(\frac{2\phi_0}{\sqrt{3} B})^{1/2}$ is the lattice constant, $\lambda$ the penetration length and $\epsilon$ the 
anisotropy parameter~\cite{blatter94}.
At zero temperature $r_p \equiv \xi$ for point impurities, with $\xi$ the 
vortex core radius. 
At high temperatures however, fast futile thermal vortex motion 
induces a growth in $r_p$, thus effectively 
smoothing the microscopic disordered potential . 
Eq.~\ref{lo_nondispersive}, which is valid for 
$L_c \geq L > \lambda^2/\epsilon a_0$ and $R_c \geq R >\lambda/\epsilon$ (we  
neglect the dispersivity in the tilt modulus~\cite{larkin79,blatter94}),  
yields, 
\begin{equation}
 R_c \approx \frac{\lambda}{\epsilon^4} \left[ \frac{l_c}{a_0} \right]^3,\;\;\;L_c \approx \frac{\lambda}{a_0} R_c,
\label{eq:larkin_lengths}
\end{equation}
for the transverse and longitudinal Larkin lengths, respectively. Pinning, 
metastability and thus irreversibility 
(and the failure of the perturbation theory) sets in at these length scales. 
Above $R_c$ and/or $L_c$ 
glassy properties are manifested.
In principle, 
reversible behaviour can be thus recovered in 
samples of dimensions $L_s \times R_s$, such 
that $R_s < R_c$ and $L_s < L_c$, provided that they still contain a large number of vortices 
$R_s/a_0 \gg 1$. Assuming that this is the situation in our samples, we can get the radius-dependent 
irreversibility line $B_i(T, R_s)$ from the equality $R_s = R_c(B_i, T)$, assuming that $L_c > L_s$. 
Using Eq.~\ref{eq:larkin_lengths} we get, 
$B_i(T, R_s) \approx  \phi_0 \left[\epsilon^4/\lambda \right]^{2/3} l_c(T)^{-2}  R_s^{2/3} 
$, where we have made the temperature dependence of the different parameters explicit 
(we neglect the temperature dependence of $\lambda$ and use its average value). 
In Fig.~3 we show that the predicted scaling $B_i \sim R_s^{2/3}$ 
produces a good collapse of the irreversibility point as a function of temperature, for different 
$R_s$. This supports our identification of the sample size $R_s$ at the onset of 
irreversibility with the Larkin radius
, although the condition $R_s > \lambda/\epsilon$ is 
not strictly satisfied for all our samples. Note also that since $R_s \gg L_s$, the 
assumption $L_c = R_c \lambda/a_0 > L_s$ is automatically satisfied, 
as $R_s/L_s > a_0/\lambda$ for our measurements.
\begin{figure}
\includegraphics[width=0.45\textwidth]{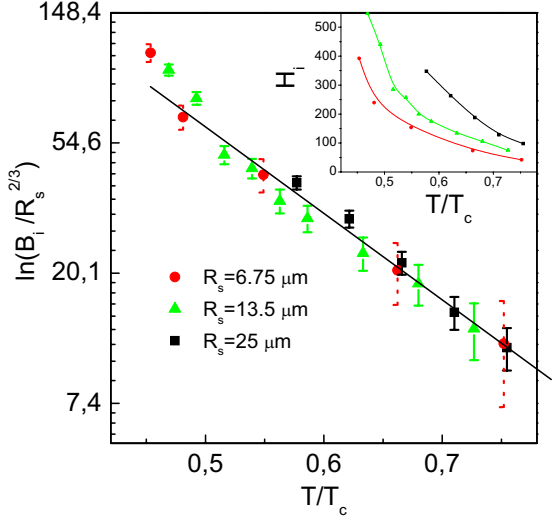}
\caption{Scaling of the irreversibility field~\cite{byh} 
through the identification of the Larkin radius with the 
sample radius 
at the onset of irreversibility. Inset: non-scaled data for samples 
of different radii.
.}
\label{fig:scaledbi}
\end{figure}
In Fig.~3 we also show that our results can be 
well described by the expression, $B_i(R_s, T) \sim R_s^{2/3} \exp(- 2T / T_0)$,
with a characteristic 
temperature $T_0 \approx 25$~K. Interestingly, Wang {\it et al.}~\cite{wang01} have reported size-effects at the
second magnetization peak in BSSCO controlled by the same exponential temperature dependence, 
with a characteristic temperature of $22.5$~K, very close to our value. 
In our calculations, the temperature dependence of $B_i$ is 
exclusively attributed to the parameter $l_c$, as $l_c(T) \sim \exp(T/T_0)$.
In order to grasp the physical meaning of this result we can assume 
that $l_c$ represents (as it indeed does in Eq.~\ref{lo_nondispersive} at zero temperature), 
the Larkin length of an isolated vortex at finite $T$. 
An exponential sensitivity $\exp[CT^{\alpha}]$ 
is consistent with the marginality of the pinning of an elastic string in a 
three-dimensional disordered medium, and it has been predicted,   
with $C$ a constant and $\alpha$ an exponent which depends on the precise nature of the disorder 
correlator function~\cite{blatter94,muller01}. In particular, 
the value $\alpha=1$ has been predicted~\cite{muller01}
for high-T$_c$ superconductors for vortex displacements $u$ satisfying 
$\xi < u < \lambda$, suitable for our case.
More interestingly, the value of $T_0$ we get is
very close to the one observed in creep~\cite{niderost96} ($\sim 20$~K), 
ac-transverse permeability~\cite{goffman98,correa01a,correa01b} ($\sim 22$~K) and critical 
current measurements~\cite{correa01a} ($\sim 20$~K) in 
samples of the same material but with radii one and two orders 
of magnitude bigger than ours, using the expected relations 
of these different quantities with $l_c$~\cite{blatter94}. 
Being $T_0 \sim U_{pc}$, with $U_{pc}$ the single pancake 
pinning energy~\cite{niderost96}, the anomalies observed near $T_0$ are 
commonly attributed to the crossover between a strong 0D pinning regime (when $l_c$ becomes 
smaller than the layer spacing and thus pancakes pin individually), 
to a weak 3D pinning regime (see Kierfield~\cite{kierfeld04} for a recent discussion).
The temperature dependence of 
$l_c$, and the fact that 
for $T>T_0$ we can use 3D weak collective pinning (Eq.~\ref{eq:larkin_lengths}) 
support our identification of the Larkin 
radius.

Our empirical estimate  
$R_c(B,T) \approx R_s \left[ B/B_i(R_s, T) \right]^{3/2} \sim  B^{3/2} \exp(T/2T_0)$,
for the weak non-dispersive pinning regime,
implies that, in the phase-space region we analyse, very big BSSCO 
samples are necessary to achieve the vortex matter thermodynamic 
limit for $T_0 < T\lesssim T_m$, with $T_m$ the 
BG melting temperature. 
This is relevant for 
the predicted crossover in the vortex lattice 
roughness, from the RM to the asymptotic BG 
regime~\cite{giamarchi95} at the characteristic scale 
$R_a = R_c (a_0/r_p)^{1/\zeta}$,
with $\zeta \sim 0.2$
the random-manifold roughness exponent~\cite{kim99,bogner01}. 
If we can roughly identify the renormalized pinning range with the 
thermally induced 
displacement, 
$r^2_p \sim \langle u^2 \rangle_{th}$~\cite{blatter94}, the Lindemann 
criterion with constant $c_L \sim 0.2$ lead us to an upper bound 
for the pinning range, $r_p^2 \leq c_L^2 a_0^2$.
We thus get $R_a \gtrsim R_c c_L^{-1/\zeta} \sim 10^3 R_c $.
If we evaluate this expression at $T$ and $B \sim  B_i(T, R_s)$, 
where we have shown that $R_c = R_s$ for our three samples, 
we find that $R_a$ is of the order of one cm. 
Our naive estimate thus indicates that the asymptotic logarithmic 
growth, characteristic of the BG phase, can be 
only achieved in huge samples in the region of the phase-space 
we analyse. This striking result seems to be however consistent with magnetic decoration experiments~\cite{kim99} 
displaying the random-manifold roughness up to distances 
$R \approx 80 a_0$, for which $W(R) \approx 0.05 a_0^2$ in the range 
$B \approx 70-120$~G. Note that the naive extrapolation of the latter 
to $R_a$, such that 
$W(R_a)=a_0^2$ gives $R_a \sim O(mm)$, in fair quantitative 
agreement with our previous estimate. These results 
indicate the remarkable possibility of detecting, in normal samples, 
the crossover from the RM to the 
BG regime at temperatures $T_0 < T$ below the irreversibility line  
as a finite-size crossover when $R_a(B,T) \approx R_s$. 
This would provide 
an independent experimental tool, 
different from neutron diffraction~\cite{klein01}, magnetic decorations
~\cite{kim99} or creep measurements~\cite{niderost96}, to test the predicted geometrical features of 
the BG phase~\cite{giamarchi95} 
in these materials.

In conclusion, we have experimentally determined, 
in a direct way, the most fundamental pinning length of 
a disordered elastic system by analysing finite-size 
crossover effects in micron-sized High-$T_c$ 
superconductors. 
This kind of study, complemented with micron-scale transport measurements can 
lead to a better understanding of the rich multi-scale physics of 
pinned vortex lattices, and of the universal properties they share 
with other pinned elastic manifolds.

We thank G. Nieva for 
providing us with 
the BSCCO raw crystals, and S. Bustingorry, V. Correa, F. de la Cruz,  D. Dom\'{\i}nguez, T. Giamarchi for very useful discussions. ABK and HP researchers of CONICET. MID fellowship holder of CNEA-CONICET.


\end{document}